\documentclass[times,twocolumn,final,authoryear]{elsarticle}

\usepackage{prletters}
\usepackage{framed,multirow}
\usepackage{natbib}
\usepackage{hyperref}
\usepackage{amssymb}
\usepackage{latexsym}

\usepackage{url}
\usepackage{xcolor}
\definecolor{newcolor}{rgb}{.8,.349,.1}

\newcommand{\quantnet}{\raisebox{-1pt}{\includegraphics[scale=0.05]{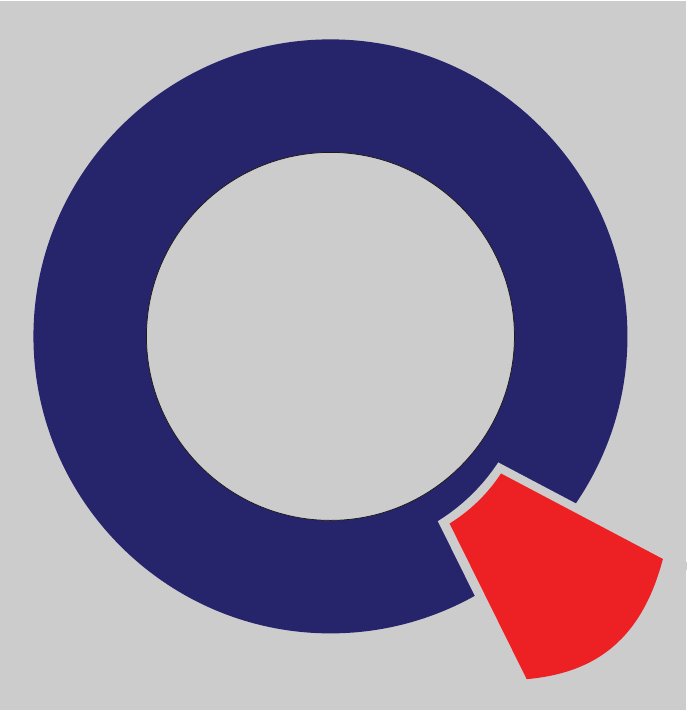}}\,}
\newcommand{\quantnetright}{\hspace*{\fill} \raisebox{-1pt}{\includegraphics[scale=0.05]{Figures/quantlet}}\,}

\journal{Pattern Recognition Letters}

\usepackage{lipsum}

\begin{document}

\thispagestyle{empty}

\begin{frontmatter}

\title{Towards the interpretation of time-varying regularization parameters in 
	streaming 
penalized regression models$^*$}

\author[1]{Lenka {Zbo\v{n}\'{a}kov\'{a}}\corref{cor1}} 
\author[2]{Ricardo Pio {Monti}}
\author[1,3,4]{Wolfgang Karl {H\"{a}rdle }}

\address[1]{C.A.S.E. - Center for Applied Statistics \& Economics, Humboldt-Universit\"{a}t zu Berlin, Spandauer Str. 1, 10178 Berlin, Germany}
\address[2]{Gatsby Computational Neuroscience Unit, UCL, 25 Howland Street, London, W1T 4JG}
\address[3]{Sim Kee Boon Institute for Financial Economics, Singapore Management University, 50 Stamford Road, Singapore 178899, Singapore}
\address[4]{The Wang Yanan Institute for Studies in Economics, Xiamen University, Xiamen, 361005 China}

\received{1 May 2013}
\finalform{10 May 2013}
\accepted{13 May 2013}
\availableonline{15 May 2013}
\communicated{S. Sarkar}

\begin{abstract}

High-dimensional, streaming datasets are ubiquitous in modern applications. 
Examples range from finance and e-commerce to the study of biomedical and neuroimaging data. 
As a result, many novel algorithms have been proposed to address challenges posed by such datasets.
In this work, we focus on the use of $\ell_1$ regularized linear models in the context of (possibly non-stationary) streaming data.
Recently, it has been noted that the choice of the regularization parameter is fundamental in such models and several methods have been proposed which iteratively tune such a parameter in a~time-varying manner; thereby allowing the underlying sparsity of estimated models to vary.
Moreover, in many applications, inference on the regularization parameter may itself be of interest, as such a parameter is related to the underlying \textit{sparsity} of the model. However,  in this work, we highlight and provide extensive empirical evidence regarding how various (often unrelated) statistical properties in the data can lead to changes in the regularization parameter. In particular, through various synthetic experiments, we demonstrate that changes in the regularization parameter may be driven by changes in the true underlying sparsity, signal-to-noise ratio or even model misspecification. The purpose of this letter is, therefore, to highlight and catalog various statistical properties which induce changes in the associated regularization parameter. 
We conclude by presenting two applications: one relating to financial data and another to neuroimaging data, where the aforementioned discussion is relevant.

\end{abstract}

%

\end{frontmatter}



\section{Introduction} 
\let\thefootnote\relax\footnote{
\hspace{-5mm}$^*$Financial support from the Deutsche Forschungsgemeinschaft via CRC 649 ``Economic Risk" , the IRTG 1792 ”High Dimensional Non Stationary Time Series”, as well as the Czech Science Foundation under grant no. 19-28231X, the Yushan Scholar Program and the European Union’s Horizon 2020 research and innovation program ''FIN-TECH: A Financial supervision and Technology compliance training programme'' under the grant agreement No 825215 (Topic: ICT-35-2018, Type of action: CSA), Humboldt-Universit\"at zu Berlin, is gratefully acknowledged.\\\textit{This is a post-peer-review, pre-copyedit version of an article published in Pattern Recognition Letters. The final authenticated version is available online at:} \url{http://dx.doi.org/10.1016/j.patrec.2019.06.021}\\ }

High-dimensional, streaming datasets pose a unique challenge to modern statisticians. 
To date, the challenges associated with high-dimensional and streaming data have been extensively studied independently. In the case of the former, a~popular avenue of research is the use of regularization methods such as the Lasso \citep{Hastie2016}. Such methods effectively address issues raised by high-dimensional data by assuming the underlying model is sparse, thereby having only a small number of non-zero coefficients. Sparse models are often easier to both estimate and interpret. Concurrently, many methods have been developed to handle streaming datasets; popular examples include sliding window methods and their generalizations to weighted moving averages \citep{Hayking2008}. 

Recently, the intersection of these two avenues of research has begun to receive increasing attention as large-scale, 
streaming datasets become commonplace. Prominent examples include \cite{Bottou2010} and \cite{Duchi2011} who propose methods through which to efficiently estimate $\ell_1$ penalized models in a streaming data context. 
However, an important aspect, which has been largely overlooked, corresponds to the optimal choice of the regularization parameter. While it is possible to employ a fixed regularization parameter, it may be the case that the statistical properties of the data vary over time, suggesting that the optimal choice of the regularization parameter may itself also vary over time. Examples of large-scale, non-stationary datasets, where the choice of the regularization parameter has been reported to be time-varying, include finance \citep{Yu2017} and neuroscience \citep{Monti2017a}. 

We note that many methods have been proposed for selecting the regularization parameter in the context of non-streaming data, the standard approach being to employ some variant of cross-validation or bootstrapping, e.g. in \citet{Hastie2016} or \citet{Chern2018}.
However, such methods are infeasible in the domain of streaming datasets due to limited computational resources.
More importantly, the statistical properties of a data stream may vary, further complicating the use of sub-sampling methods.
Recently, methods to handle time-varying regularization parameters have been proposed.
\cite{Monti2018sadm} propose a novel framework through which to iteratively infer a time-varying regularization parameter via the use of adaptive filtering. The proposed framework is developed for penalized linear regression (i.e., the Lasso) and subsequently extended to penalized generalized linear models. \cite{zbonakova2017time} study the dynamics of the regularization parameter, focusing particularly on quantile regression in the context of financial data. Using sliding windows method, they demonstrate that the choice of time-varying regularization parameter based on the adjusted Bayesian information criterion (BIC) is closely correlated with the financial volatility. 
The BIC was employed, as such a choice of parameter is optimal in terms of model consistency.

While the aforementioned methods correspond to valuable contributions, the purpose of this paper is to highlight potential shortcomings when interpreting time-varying regularization parameters. In particular, we enumerate several (often unrelated) statistical properties of the underlying data which may lead to changes in the optimal choice of the regularization parameter. This paper, therefore, serves to highlight important issues associated with the interpretation of time-varying regularization parameters as well as the associated model parameters. 

The remainder of this paper is organized as follows.
We formally outline the challenge of tuning time-varying regularization parameters as well as related work in Section \ref{sec::PreLim}.
In Section \ref{sec::ExpRes}, we present extensive empirical results, highlighting how various aspects of the underlying data may result in changes in the
estimated regularization parameter. Computations included in this work were performed with the help of R software environment \citep{rcore2014} and we provide code to reproduce all experiments at \href{http://quantlet.de}{\protect\quantnet{Quantlet}} platform.

\section{Preliminaries and related work}\label{sec::PreLim}

In this work, we focus on streaming linear regression problems. Formally, it is assumed that we observe a sequence of pairs
$(X_t, y_t)$, where $X_t \in \mathbb{R}^p$ corresponds to a $p$-dimensional vector of predictor variables and $y_t \in \mathbb{R}$ is a univariate response. The objective of penalized streaming linear regression problems consists in 
accurately predicting future responses, $y_{t+1}$, from predictors $X_{t+1}$ via a linear model. 
Following the work of \cite{tibshirani1996regression}, an $\ell_1$ penalty, parameterized by $\lambda \in \mathbb{R}_+$,  is subsequently introduced in order to encourage sparse solutions as well as ensure the associated optimization problem is well-posed. 
For a~pre-specified choice of fixed regularization parameter, $\lambda$, time-varying regression coefficients can be estimated by minimizing the following convex objective:
\begin{equation}
L_t (\beta, \lambda) = \sum_{i=1}^t w_i \left ( y_i - X_i^{\top} \beta \right )^2 + \lambda ||\beta||_1,
\label{ConvexObjectiveLasso}
\end{equation}
where $w_i >0$ are weights indicating the importance given to past observations \citep{aggarwal2007data}. 
For example, it is natural to allow $w_i$ to decay monotonically in a manner which is proportional to the chronological proximity of the 
$i$th observation. While the weights $w_i$ may be tuned using a fixed forgetting factor, throughout this work we opt for 
the use of a sliding window due to the simplicity of the latter method. 

In the context of non-stationary data the optimal estimates of regression coefficients, $\hat \beta_t$, may vary over time and several methods have been proposed in order to address this issue \citep{Bottou2010, Duchi2011}. However, the same argument can be posed in terms of the associated regularization parameter, $\lambda$. The choice of such a parameter dictates the severity of the associated $\ell_1$ 
penalty, implying that different choices of $\lambda$ will result in vastly different estimated models. While there exists a large range of methodologies through which to iteratively update the regression coefficients, the choice of the regularization parameter has, until recently,  been largely overlooked. 
Lately, \cite{Monti2018sadm} proposed a framework through which to 
learn time-varying regularization parameter in a streaming scenario.
The proposed framework is motivated by adaptive filtering theory \citep{Hayking2008} and seeks to iteratively update the regularization parameter via 
stochastic gradient descent.
In related work, \cite{zbonakova2017time} focus on the choice of the regularization parameter in the context of a quantile regression model. They 
propose the use of sliding windows and information theoretic quantities 
to select the associated regularization parameter. 

Formally, \cite{Osborne2000lassodual} clearly outline the relationship between the Lasso parameter, $\lambda$, and the 
data. They note that the regularization parameter may be interpreted as the Lagrange multiplier associated with a constraint on the 
$\ell_1$ norm of the regression coefficients. As such, considering the dual formulation yields:
\begin{equation}
\lambda = \frac{\lbrace \textbf{Y} - \textbf{X}\hat{\beta}(\lambda)\rbrace ^{\top}\textbf{X}\hat{\beta}(\lambda)}{||\hat{\beta}(\lambda)||_1},
\label{eq:lambda}
\end{equation}
where we have ignored the weights, $w_i$, and use bold notation to denote vectors and matrices respectively. 
Note that in equation (\ref{eq:lambda}) we clearly denote the dependence of the estimated regression coefficients on 
$\lambda$. 
As a result, we observe three main effects driving the optimal choice of the regularization parameter:
\begin{enumerate}
	\item Variance or magnitude of the residuals: $ \textbf{Y}- \textbf{X} \hat \beta(\lambda)$. 
	As the variance of residuals increases so does the associated regularization parameter, leading to 
	an increase in sparsity of $\hat \beta(\lambda)$. This is 
	natural as an increase of the variance of residuals is indicative of a drop in the signal-to-noise ratio of the data. 
	\item The $\ell_1$ or $\ell_0$ norm of the model coefficients: $||\hat{\beta}(\lambda)||_1$. As this term appears in the denominator of equation (\ref{eq:lambda}), it is inversely correlated with the regularization parameter. 
	This is to be expected as we require a small
	regularization parameter in order to accurately recover regression coefficients with large $\ell_1$ norm.
	\item Covariance structure of the design matrix: $\textbf{X}$. The term related to the covariance structure of the design matrix, $\textbf{X}^{\top}\textbf{X}$, can be extracted from the elements in the numerator of equation
	 (\ref{eq:lambda}). This suggests that the covariance matrix of the predictors will have a significant impact on the value of the regularization parameter, $\lambda$. 
	 We note that this effect will also affect the $\ell_1$ and $\ell_0$ norms of the model coefficients, resulting in 
	 a complicated relationship with the regularization parameter. In Section \ref{Sim_rho} we demonstrate the non-linear nature of this relationship.
\end{enumerate}

As such, it follows that multiple aspects of the data may influence the choice of the associated regularization parameter.
Crucially, whilst such a parameter is often interpreted as being indicative of the \textit{sparsity} of the underlying model, 
equation (\ref{eq:lambda}) together with the aforementioned discussion demonstrates that this is not necessarily the case. In the remainder of this work, we provide extensive empirical evidence to validate  these claims.

\section{Experimental results}

In this section, we provide an extensive simulation study to demonstrate the 
effects of the three aforementioned model properties on the choice of the optimal regularization parameter. 
Based on the observations from Section \ref{sec::PreLim}, we designed a series of experiments where 
one property of the data was allowed to vary whilst the remaining two were left unchanged. 
A further concern is to show that if two or more of the properties of the data should simultaneously change it can result in cancelling out their effects on 
the regularization parameter. Further experiments were designed to study those scenarios. 
The purpose of the experimental results presented in this section is two-fold. 
First, we identify the various statistical properties which cause the optimal choice of regularization parameter to vary.
Second, we also highlight how changes of such properties interact with each other and catalog their joint effects on the 
choice of the regularization parameter.

\label{sec::ExpRes}
\subsection{Synthetic data generation}

We focus exclusively on a linear model of the form:
$$y_t = X_t\beta_t + \varepsilon_t.$$ We define the 
 number of observations as $n$, the number of non-zero parameters as $q = ||\beta||_0 \leq p$ and an \textit{iid} error term $\varepsilon = (\varepsilon_1, \ldots, \varepsilon_n)^{\top}$, such that $\varepsilon_t \sim (0, \sigma_t^2)$. The $p$-dimensional vector of predictor variables $X_t$ was generated from the normal distribution $\mbox{N}_p(0, \Sigma)$, where the elements of $(p \times p)$ covariance matrix $\Sigma = (\sigma_{ij})_{i,j = 1}^p$ were set to be $\sigma_{ij} = \rho^{|i - j|}$, for $i, j = 1, \ldots, p$, with a correlation parameter $\rho$.
We generate synthetic data where one of the following  properties varies over time (thereby resulting in non-stationarity):
\begin{enumerate}
	\item Time-varying variance of residuals: $\sigma_t^2$ varies over time.
	\item Time-varying  $\ell_1$ or $\ell_0$ norm of regression coefficients: $q$ varies over time. 
	\item Time-varying correlation within design matrix: $\rho$ varies over time. 
\end{enumerate}

For each experiment, the total number of observations was set to $n=400$ with a dimensionality of $p=20$.
The optimal choice of the regularization parameter (together with associated regression coefficients) was estimated 
using three distinct methods.
We consider the use of the sliding window method in combination with both 
Bayesian information criterion (BIC) and generalized cross-validation (GCV) 
to select the associated regularization parameter. Finally, the gradient method proposed by 
\cite{Monti2018sadm}, named Real-time Adaptive Penalization (RAP), is also applied.  A burn-in period of 50 observations was employed to obtain an initial estimate for regression coefficients as well as $\lambda$.  Each experiment was repeated $100$ times and the mean value of the regularization parameter was studied.

\subsubsection{Change of the variance of residuals}

We begin by studying the effect of residual variance on the 
choice of the regularization parameter, $\lambda$. 
The regression coefficients were set to 
$\beta_t = (1, 1, 1, 1, 1, 0, \ldots, 0)^{\top}$,  yielding $q = 5$ and the covariance parameter was set to be $\rho = 0.5$. 
The vector of residuals was simulated according to a piece-wise stationary distribution as follows:
\begin{equation}
\varepsilon_t \sim \left\lbrace \begin{array}{lcl}
\medskip 
\mbox{N}(0, \sigma^2_1), & \mbox{for} & t < 200;\\
\mbox{N}(0, \sigma^2_2), &       & t \geq 200,
\end{array} \right.
\label{eq:resvector}
\end{equation}
resulting in a significant change in the variance of residuals at the 200th observation. 
Throughout these experiments, we set $\sigma_1=1$ and allowed $\sigma_2$ to vary from $\sigma_2  \in \{1.1, \ldots, 2\}$.
In order to study the effects of changes in the variance of residuals, we consider the 
change in the estimated regularization parameter defined by the ratio of the values of $\lambda$ after ($\lambda_2$) and before ($\lambda_1$) the change point as a function of the ratio $\sigma_2/ \sigma_1$. Following the 
discussion from Section \ref{sec::PreLim}, we would expect larger values of the ratio 
to yield larger changes in the choice of the regularization parameter.

\begin{figure}[h!]
\centering
\includegraphics[scale=0.6]{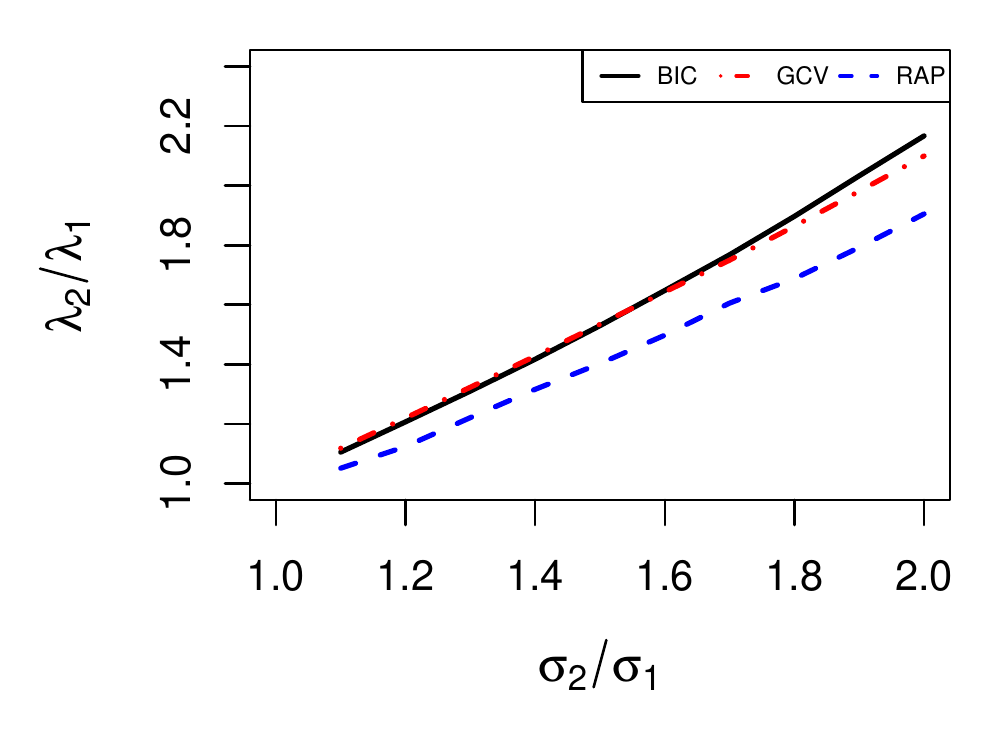}
\vspace{-3mm}
\caption{Relative changes of $\lambda$ in dependence on relative changes of the standard deviation $\sigma$.}
\label{fig:deltasigma}
\href{https://github.com/QuantLet/TimeVaryingPenalization/tree/master/TVRPchangeSQR}{\protect\quantnetright{TVRPchangeSQR}}
\end{figure}

Figure \ref{fig:deltasigma} plots the effect of the changes in the standard deviation of residuals on the Lasso parameter $\lambda$. As expected when looking at the formula (\ref{eq:lambda}), there is a linear dependency visible. In case of the BIC and GCV as selection criteria for the values of $\lambda$, the line is almost identical. For the RAP algorithm, $\lambda$ changes slower, but the effect can be clearly seen.

\begin{figure}[h!]
\centering
\includegraphics[scale=0.5]{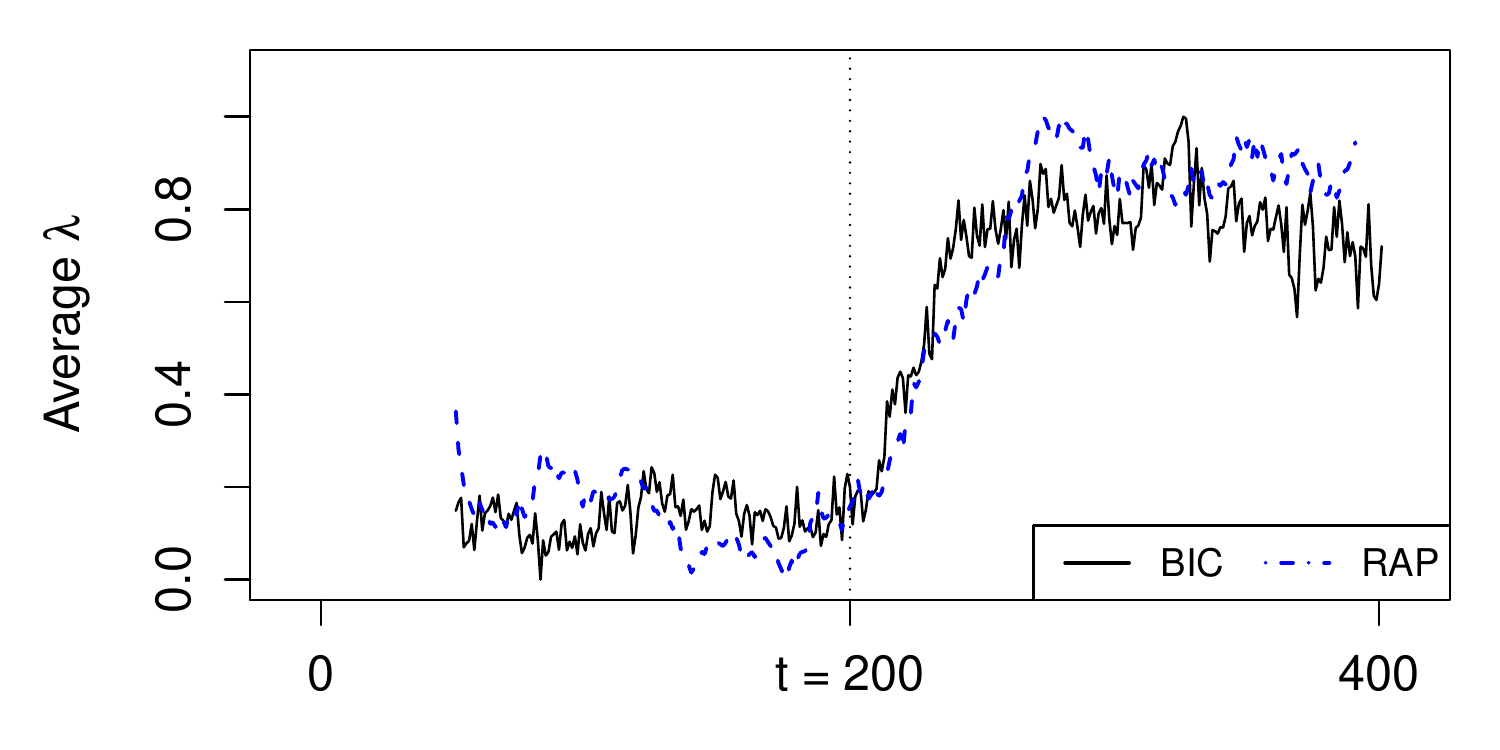}
\vspace{-3mm}
\caption{Standardized series of average $\lambda$ over 100 scenarios with a change point at $t = 200$ and $\sigma_1 = 1$ and $\sigma_2 = 1.5$. 
}
\label{fig:deltasigmaseries}
\href{https://github.com/QuantLet/TimeVaryingPenalization/tree/master/TVRPchangeSQR}{\protect\quantnetright{TVRPchangeSQR}}
\end{figure}

In order to illustrate how the series of values of the Lasso parameter changes over time and how long it takes to adjust for the new settings of the model, we depict the average $\lambda$ over the 100 scenarios in Figure \ref{fig:deltasigmaseries}, where $\sigma_1 = 1$ and $\sigma_2 = 1.5$. Since the BIC and GCV yield very similar results, we omit the GCV in this case and normalize the BIC and RAP values of $\lambda$ to fit into the interval [0, 1].

From Figure \ref{fig:deltasigmaseries} it is clear that the values of $\lambda$ adjust for the new model settings for the whole length of the moving window (50 in this case) if the BIC is implemented and for the RAP algorithm the adjustment is dependent on the size of the fixed forgetting factor. $r$. 
We note there is a clear change in the regularization parameter following $t=200$, indicating the need to
adaptively estimate the regularization parameter and demonstrating the drawback of using a fixed and pre-specified value of $\lambda$.

\subsubsection{Change of the $\ell_1$ and $\ell_0$ norm of $\beta$}

It follows that the choice of the regularization parameter is closely related to the 
true underlying $\ell_1$ and $\ell_0$ norm of the regression coefficients; the relation to the latter is because 
the Lasso constraint is introduced as a convex relaxation of the $\ell_0$ norm. 

\begin{figure}[b!]
\centering
\includegraphics[scale=0.5]{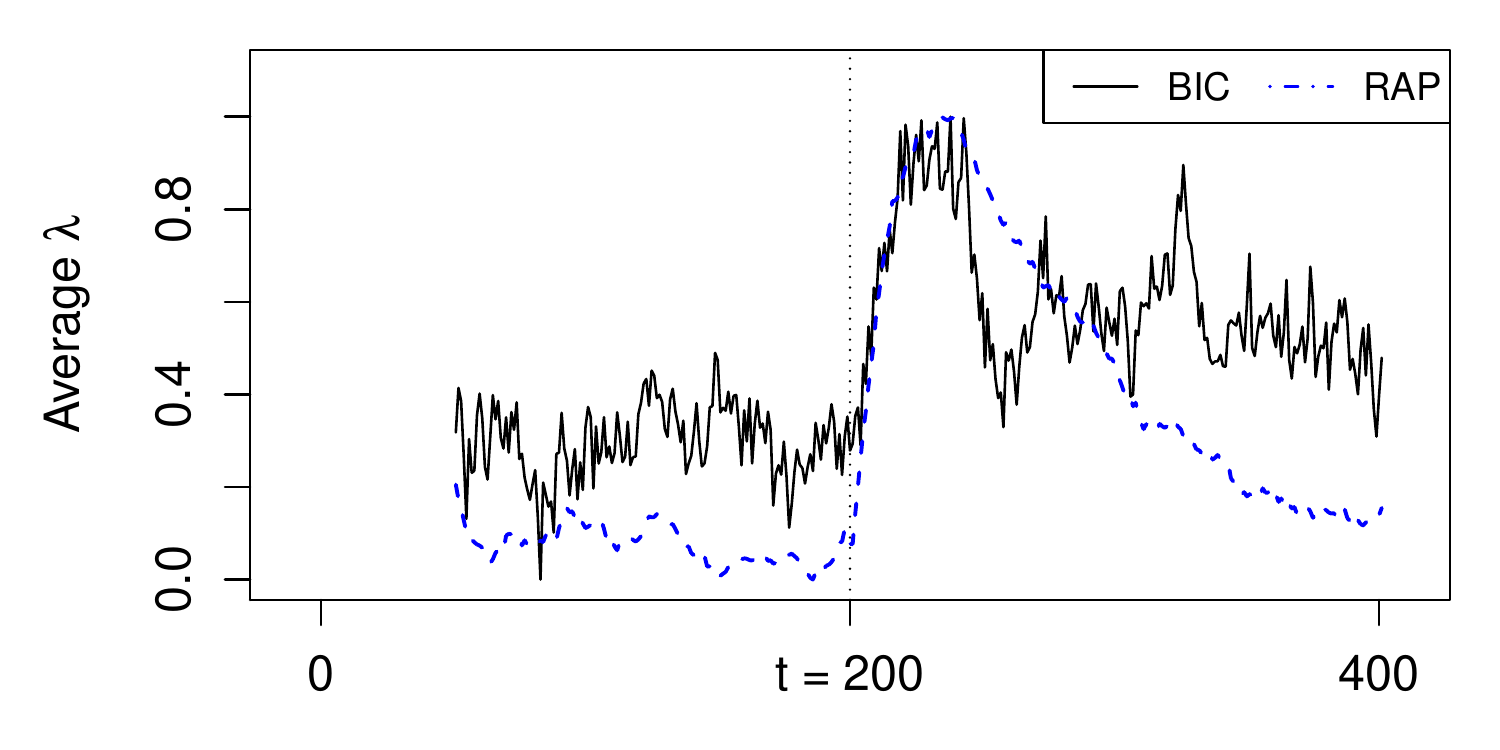}
\vspace{-3mm}
\caption{Standardized series of average $\lambda$ over 100 scenarios with a change point at $t = 200$ and regression coefficients $\beta_t$ defined by (\ref{eq:l1normbeta}). 
}
\label{fig:deltal1series}
\href{https://github.com/QuantLet/TimeVaryingPenalization/tree/master/TVRPchangeB}{\protect\quantnetright{TVRPchangeB}}
\end{figure}

In this set of simulations, we therefore quantify the effects of changes in both the $\ell_0$ and $\ell_1$ norms on the 
optimal choice of the regularization parameter. In particular, 
we set $\sigma_1 = \sigma_2 = 1$ and $\rho = 0.5$. 
As a first example, we consider the following changes in the $\ell_1$ norm:
\begin{equation}
\beta_t = \left\lbrace \begin{array}{lcl}
\medskip 
(1, 1, 1, 1, 1, 0, \ldots, 0)^{\top}, & \mbox{for} & t < 200;\\
(1, 0.8, 0.6, 0.4, 0.2, 0, \ldots, 0)^{\top}, &       & t \geq 200.
\end{array} \right.
\label{eq:l1normbeta}
\end{equation}
The time series of estimated $\lambda$ values is presented in Figure \ref{fig:deltal1series}.

We note that the change in the $\ell_1$ norm of the model coefficients $\beta$ results in an upward trend in $\lambda$ for the BIC parameter choice which is visible in the long run. For a short period after the change, exactly the period of 50 observations from the moving window, the misspecification of the model drives the size of residuals and with them, the values of $\lambda$ higher and lower again in a ``bump"-shaped line. The same holds for the RAP algorithm, however, because of the fixed forgetting factor, the values of $\lambda$ are adjusting to the new model settings more slowly. 

In order to study the effect of changes in the $\ell_0$ norm (i.e., the size of the active set) we generated synthetic data whereby:
 \begin{equation}
||\beta_t||_0 = \left\lbrace \begin{array}{lcl}
\medskip 
q_1, & \mbox{for} & t < 200;\\
q_2, &       & t \geq 200,
\end{array} \right.
\label{eq:l0normbeta}
\end{equation}
with $q_1  = 5$ and $q_2 \in \{ 6, \ldots, 10, 15 \}$. 

\begin{figure}[h!]
\centering
\includegraphics[scale=0.6]{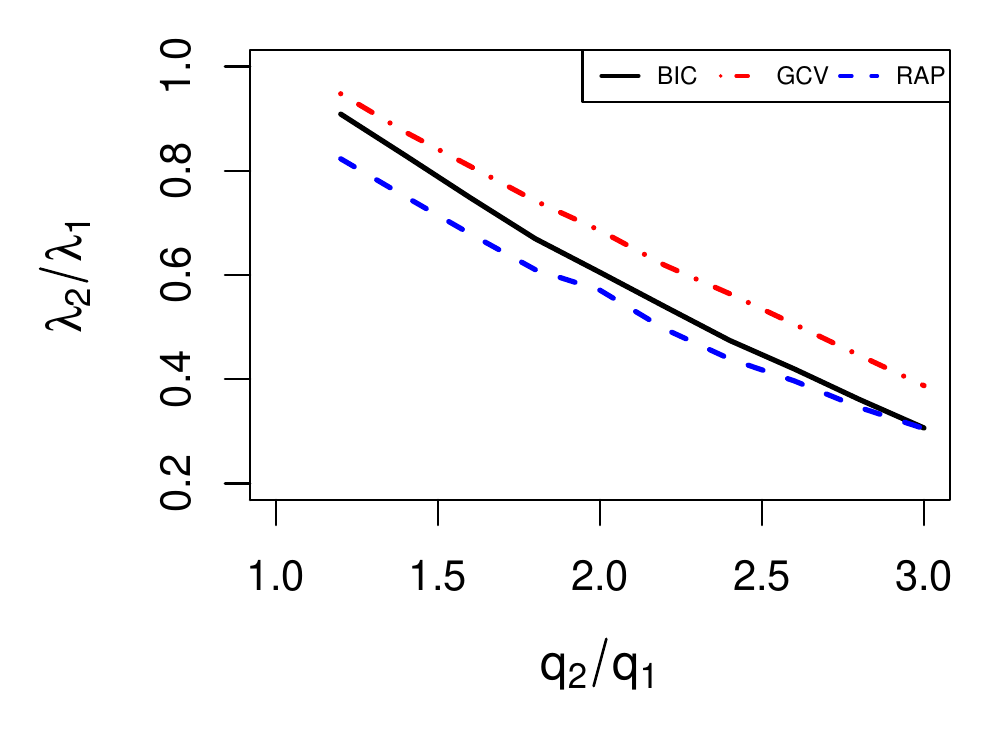}
\vspace{-3mm}
\caption{Relative changes of $\lambda$ in dependence on relative changes of the size of the active set $q$. 
}
\label{fig:deltaq}
\href{https://github.com/QuantLet/TimeVaryingPenalization/tree/master/TVRPchangeSQR}{\protect\quantnetright{TVRPchangeSQR}}
\end{figure}

Figure \ref{fig:deltaq} visualizes the relative changes of $\lambda$ as a function of the relative changes 
in the size of the active set, defined as $q_2/q_1$. 
We note there is a clear decay of values of $\lambda$ as $q_2/q_1$ increases. This is to be expected, as an increase in 
the specified ratio implies a larger active set in the latter part of the time series.  
This figure provides empirical validation of the inverse correlation between the 
magnitude of the active set and the estimated regularization parameter. 

\subsubsection{Change of covariance parameter $\rho$}
\label{Sim_rho}
Finally, we study the effect of changes in the covariance structure of features, $X_t$, 
on the regularization parameter. We note that whilst it is possible to vary the covariance structure in many ways, 
we consider a simple model whereby $\Sigma = (\sigma_{ij})_{i,j=1}^p$
and set $\sigma_{ij} = \rho^{|i-j|}$. The benefit of such a model is that it only depends on a single parameter, $\rho$, 
simplifying the interpretation and visualization of results. 
As such, we investigate changes in the covariance parameter $\rho$, while fixing $\sigma=1$ and $q=5$. 
Formally, piece-wise stationary data was generated such that:
  \begin{equation}
 \rho_t= \left\lbrace \begin{array}{lcl}
 \medskip 
 \rho_1, & \mbox{for} & t < 200;\\
 \rho_2, &       & t \geq 200,
 \end{array} \right.
 \label{eq:l0normbeta}
 \end{equation}
where
$\rho_1 = 0.1$ and  $\rho_2 \in \{ 0.2, 0.3, \ldots, 0.9\} $. 

\begin{figure}[h!]
\centering
\includegraphics[scale=0.6]{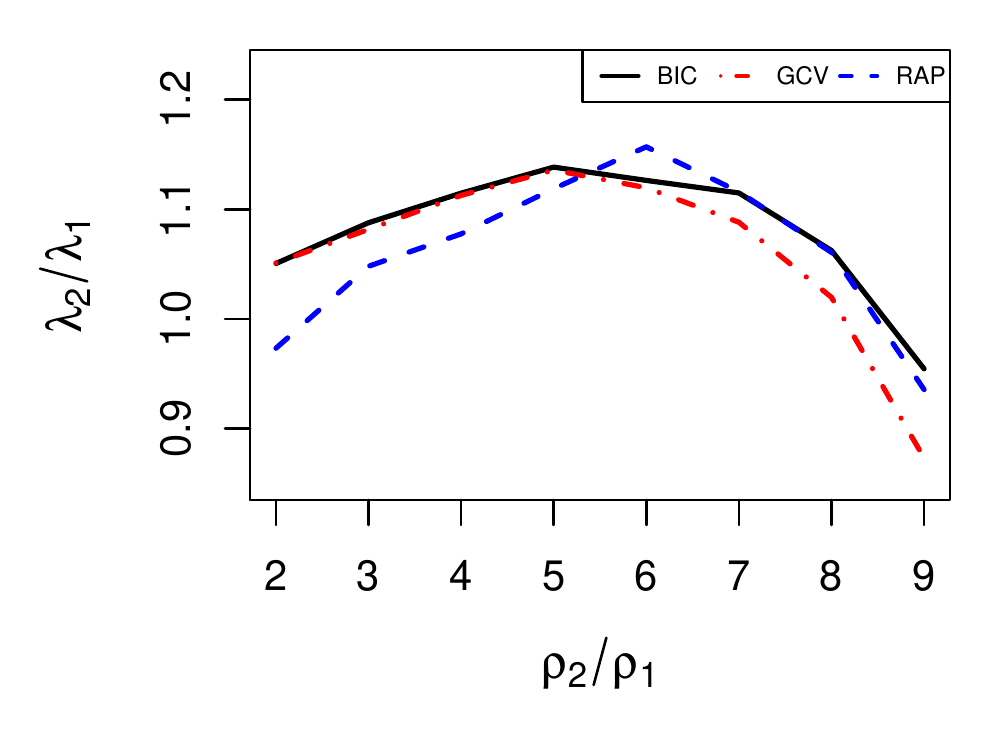}
\vspace{-3mm}
\caption{Relative changes of $\lambda$ in dependence on relative changes of the correlation parameter $\rho$.  
}
\label{fig:deltarho}
\href{https://github.com/QuantLet/TimeVaryingPenalization/tree/master/TVRPchangeSQR}{\protect\quantnetright{TVRPchangeSQR}}
\end{figure}

\begin{figure}[b!]
\centering
\includegraphics[scale=0.45]{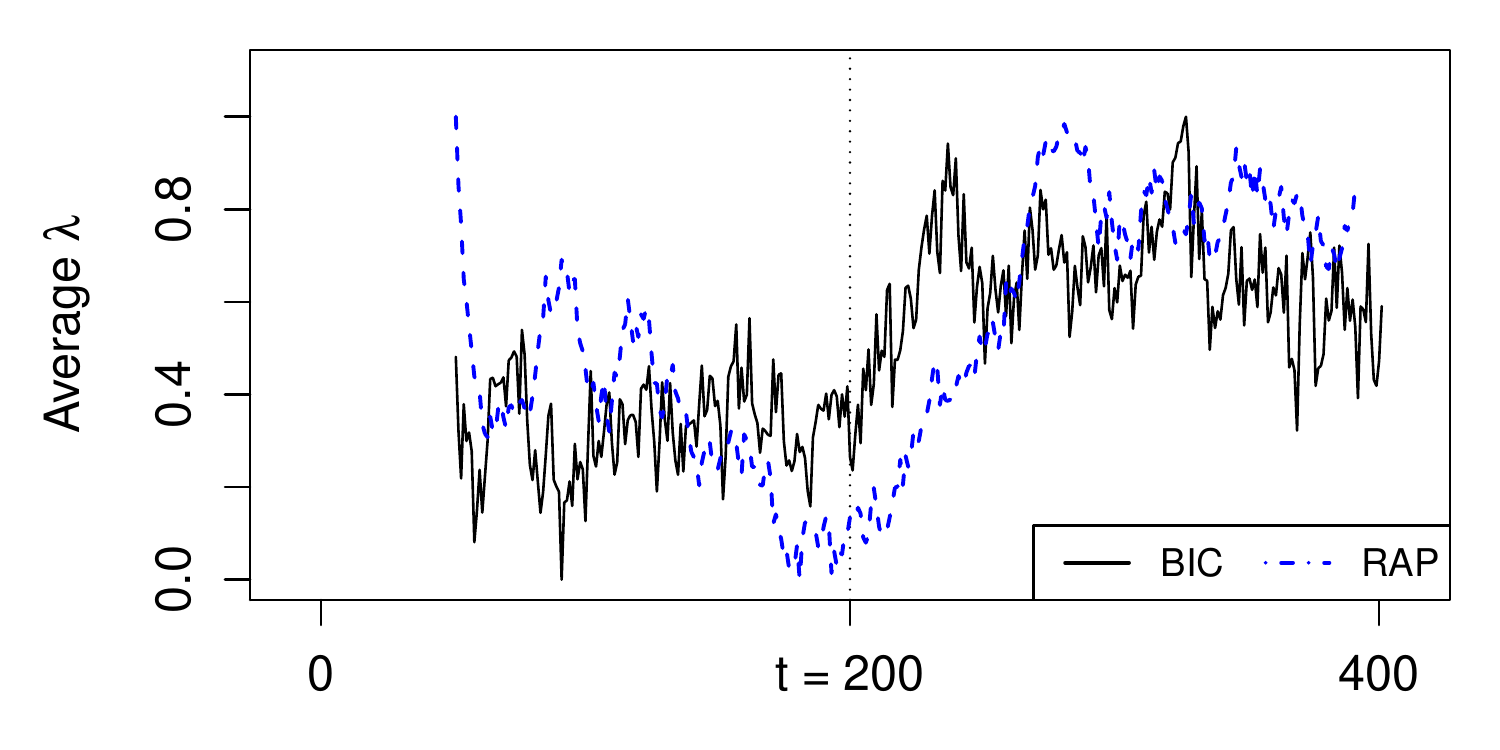}
\vspace{-3mm}
\caption{Standardized series of average $\lambda$ over 100 scenarios with a change point at $t = 200$ and $\rho_1 = 0.1$ and $\rho_2 = 0.5$. 
}
\label{fig:deltarhoseries}
\href{https://github.com/QuantLet/TimeVaryingPenalization/tree/master/TVRPchangeSQR}{\protect\quantnetright{TVRPchangeSQR}}
\end{figure}

As in the previous experiments, we visualize the relative changes of $\lambda$ with respect to the relative changes of $\rho$ in Figure \ref{fig:deltarho}. The time series of the estimated values of $\lambda$ over the whole sample size are depicted in Figure \ref{fig:deltarhoseries}.

From Figure \ref{fig:deltarho}
it is important to note that the changes of $\lambda$ no longer demonstrate a linear dependency with the statistical property of interest. 
For $\rho_2 = 0.2, \ldots, 0.8$ the values of $\lambda$ tend to rise with a rising covariance of the predictors and the biggest change occurs for $\rho_2 = 0.5$ in the case of the BIC and GCV. In the RAP method example, the values of $\lambda$ decrease for $\rho_2 = 0.2$ and $0.9$ and the biggest change is visible in the case that $\rho$ changes to the value $\rho_2 = 0.6$. 

A potential explanation for the non-linear nature of the relationship demonstrated in Figure \ref{fig:deltarho} 
is due to the selection properties of the Lasso.
It is widely acknowledged that in the presence of strongly correlated variables (corresponding to large $\rho$ values) the Lasso tends to choose only a single variable form the group of  strongly correlated covariates (indeed this phenomenon is the inspiration for the elastic net \citep{zou2005regularization}). As such, as $\rho$ increases, the term $\textbf{X}^{\top}\textbf{X}$ from the numerator of $\lambda$ drives its values higher. If the $\rho$ value is too high, we speak of multicollinearity, where the denominator of $\lambda$ is affected and becomes larger, which consequently causes the $\lambda$ values to drop.

In Figure \ref{fig:deltarhoseries} the change from $\rho_1 = 0.1$ to $\rho_2 = 0.5$ is depicted. 
We note there is a change in $\lambda$ despite the fact that the $\ell_1$ and $\ell_0$ norms remain unchanged.

\subsubsection{Simultaneous changes of model specifications}

While the previous experiments have examined the effects of changing a single property of the data, we 
now consider combinations of specific changes. In particular, the purpose of the remaining experiments is to
highlight how simultaneous changes to two properties of the data result in a \textit{canceling out}
the effects on the regularization parameter.
The purpose of this section is, therefore, to highlight 
the fact that it is possible to have a non-stationary data where the three properties discussed in 
Section \ref{sec::PreLim} vary, and yet the optimal choice of the sparsity parameter is itself constant.

\begin{figure}[b!]
\centering
\includegraphics[scale=0.45]{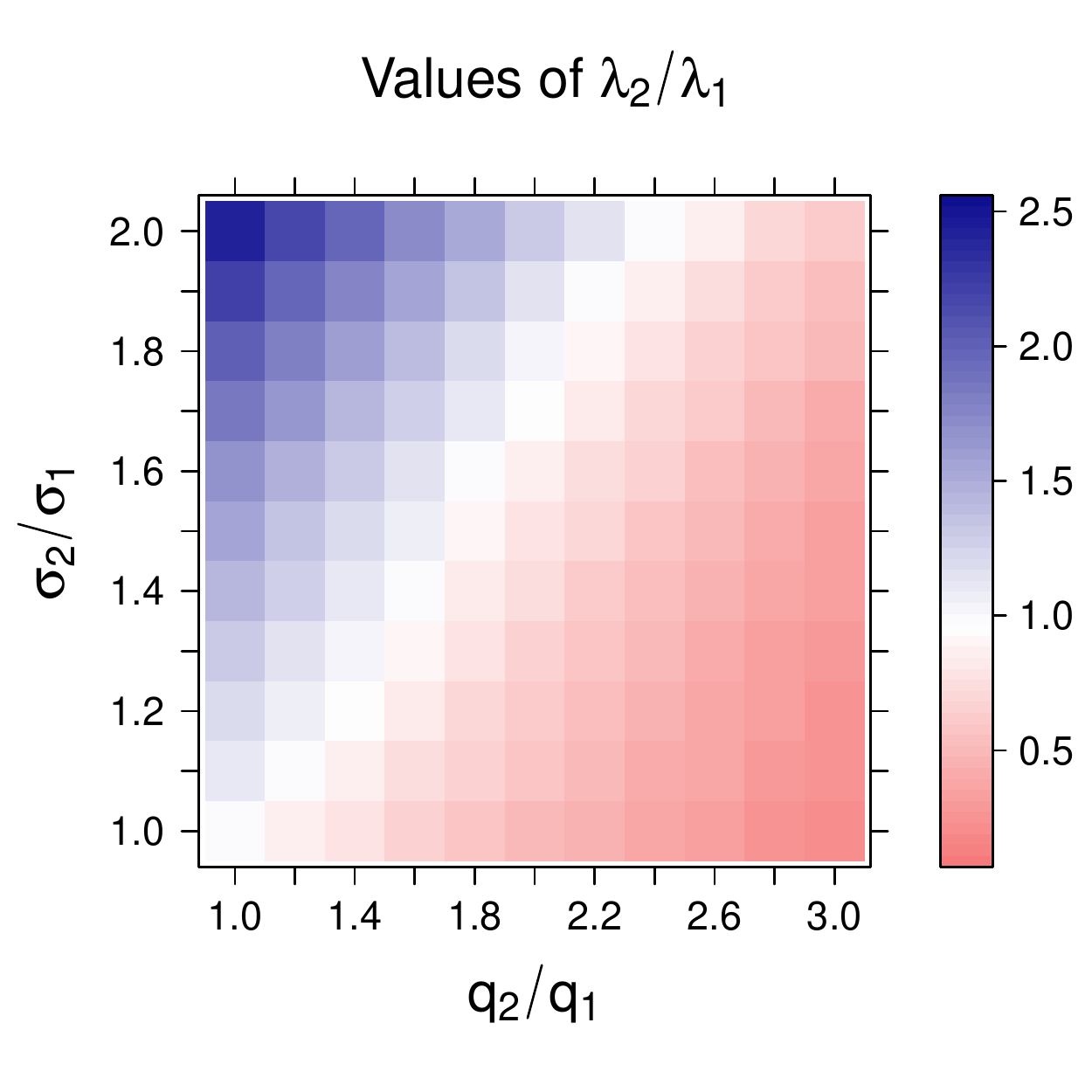}
\vspace{-3mm}
\caption{Relative changes of $\lambda$ corresponding to the combination of relative changes of $q$ and $\sigma$.}
\label{fig:QvsSigma}
\href{https://github.com/QuantLet/TimeVaryingPenalization/tree/master/TVRPchangeSQR}{\protect\quantnetright{TVRPchangeSQR}}
\end{figure}

We begin by studying simultaneous changes in the $\ell_0$ or $\ell_1$ norm of regression parameters, $\beta_t$, together with 
 changes in the variance of residuals, $\sigma^2$. 
Recall that the optimal choice of regularization parameter was positively correlated with the 
magnitude of residuals (see Figure \ref{fig:deltasigma})
whilst being negative correlated with $q$ (see Figure \ref{fig:deltaq}).
 Figure \ref{fig:QvsSigma} shows the relative change of $\lambda$ as a~function of both 
 $q_2/q_1$ and $\sigma_2/\sigma_1$. It is important to note the diagonal trend, 
 which indicates that for any increase in $q$, a proportional increase in $\sigma$ directly 
 cancels out the change in the estimated regularization parameter. 
 This is a natural result, as the changes in $\sigma$ influence the numerator, whilst changes in the $\ell_0$ or $\ell_1$ norm affect the denominator in (\ref{eq:lambda}).

\begin{figure}[t!]
	\centering
	\includegraphics[scale=0.5]{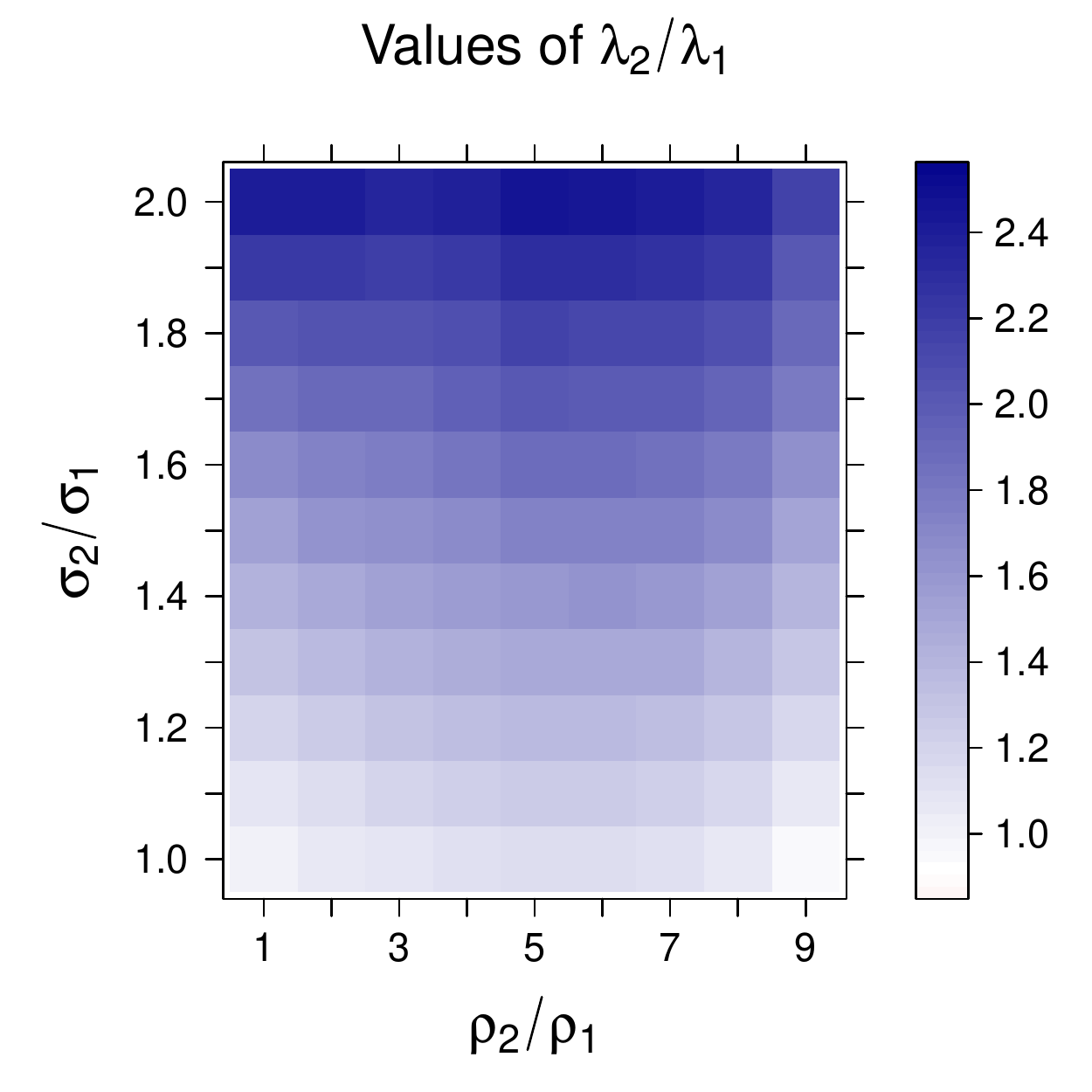}
	\vspace{-3mm}
	\caption{Relative changes of $\lambda$ corresponding to the combination of relative changes of $\rho$ and $\sigma$.}
	\label{fig:RhovsSigma}
	\href{https://github.com/QuantLet/TimeVaryingPenalization/tree/master/TVRPchangeSQR}{\protect\quantnetright{TVRPchangeSQR}}
\end{figure}

Next, we consider the combination of varying the covariance parameter, $\rho$,  and  the 
variance of the residuals, parameterized by $\sigma$.
Recall from the previous discussion that the covariance parameter, $\rho$, did not have a linear relationship with the regularization parameter, $\lambda$. 
Such a non-linear relationship can be clearly seen again in 
Figure \ref{fig:RhovsSigma}. 
Furthermore, we note that changes in $\sigma$ tend to dominate the changes in $\rho$ with the 
largest changes in $\lambda$ occurring  for large changes in $\sigma$.

\begin{figure}[b!]
\centering
\includegraphics[scale=0.5]{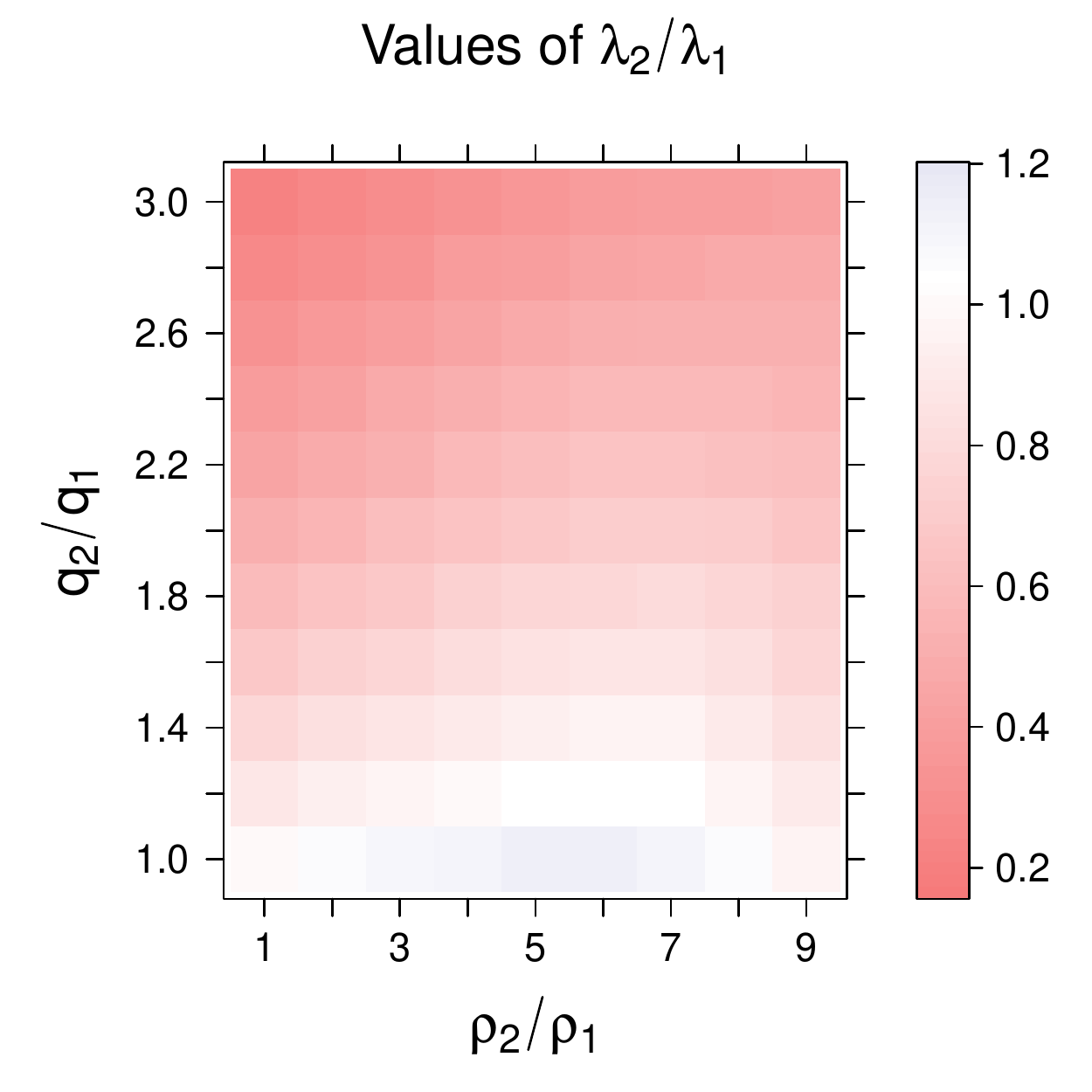}
\vspace{-3mm}
\caption{Relative changes of $\lambda$ corresponding to the combination of relative changes of $q$ and $\rho$.}
\label{fig:QvsRho}
\href{https://github.com/QuantLet/TimeVaryingPenalization/tree/master/TVRPchangeSQR}{\protect\quantnetright{TVRPchangeSQR}}
\end{figure}

Finally, we also studied the combination of changes in the 
$\ell_0$ norm, denoted by $q$, together with changes in the covariance parameter, $\rho$. 
Note that changes in these parameters are strongly coupled due to the effect of multicollinearity induced by simultaneously 
increasing the number of non-zero regression coefficients together with their correlations.
The results, provided in Figure \ref{fig:QvsRho}, highlight these dependencies. 
For the values of $\rho$ near $\rho = 0.5$, there are some combinations which cancel each other. For the extreme parts of the heatmap, e.g. $\rho_2 = 0.2$ or $\rho_2 = 0.9$, the pattern is clearly driven by the change in the active set only.

\subsection{Application to financial and neuroimaging data}

Until now we have provided extensive 
empirical evidence based on a variety of simulations, each
varying one or more of the statistical properties of the data. 
In this section, we conclude by presenting two distinct real-world 
datasets were we observe significant variability  
in the time-varying regularization parameter.  
The two examples presented in this section provide a clear illustration that non-stationary 
data are present in a wide range of applications.

We study two high-dimensional real-world datasets from distinct applications: 
the first consists of stock returns and the second corresponds to 
functional MRI (fMRI) dataset taken from an emotion task.
The stock return data consists of daily stock returns of 100 largest financial companies over a period of 11 years from 2007 to 2018. The companies listed on NASDAQ are ordered by the market capitalization and downloaded from Yahoo Finance. This data is particularly interesting as it covers the financial crisis of 2008-2009. By analyzing this data, it is hoped that we may be able to
understand the statistical properties which directly precede similar financial crises, thereby potentially 
providing some form of advanced warning.

The second dataset we consider corresponds to fMRI data 
collected as a part of the Human Connectome Project (HCP). This dataset 
consists of measurements of 15 distinct brain regions
taken during an Emotion task, as described in \cite{barch2013function}. Data was analyzed over a subset of 
50 subjects. 
While traditional neuroimaging studies 
were premised on the assumption of stationarity, an exciting avenue of neuroscientific research corresponds to 
understanding the non-stationary properties of the data and how these may potentially 
correspond to changes induced by distinct tasks \citep{Monti2017a} or changes across subjects \citep{Monti2017}.

\begin{figure}[t!]
\centering
\includegraphics[scale=0.5]{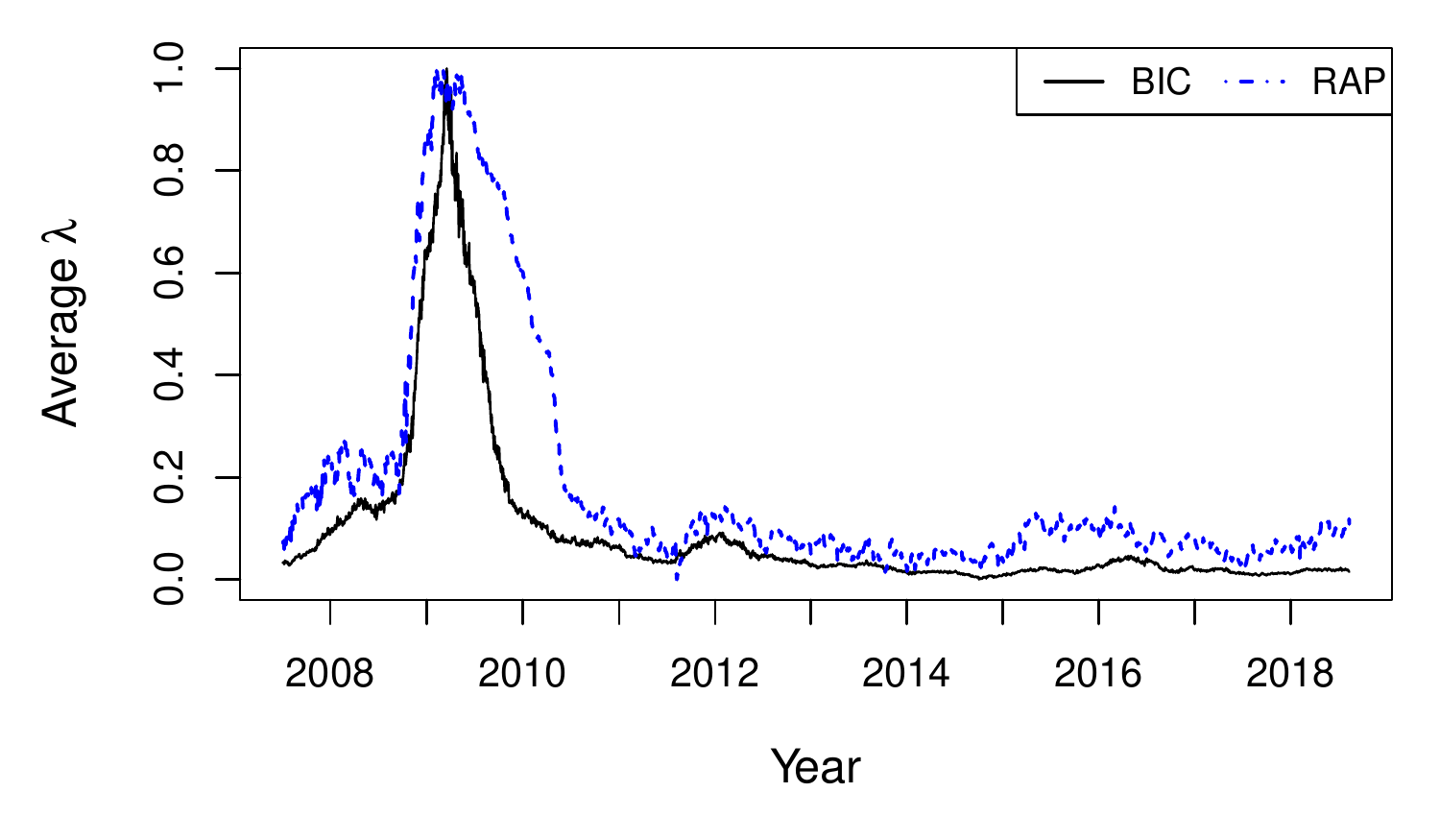}
\vspace{-3mm}
\caption{Standardized series of average $\lambda$ in the US stock returns data, daily observations from January 3, 2007 to August 10, 2018.
}
\label{fig:finance}
\href{https://github.com/QuantLet/TimeVaryingPenalization/tree/master/TVRPfrm}{\protect\quantnetright{TVRPfrm}}
\end{figure}

\begin{figure}[t!]
\centering
\includegraphics[scale=0.45]{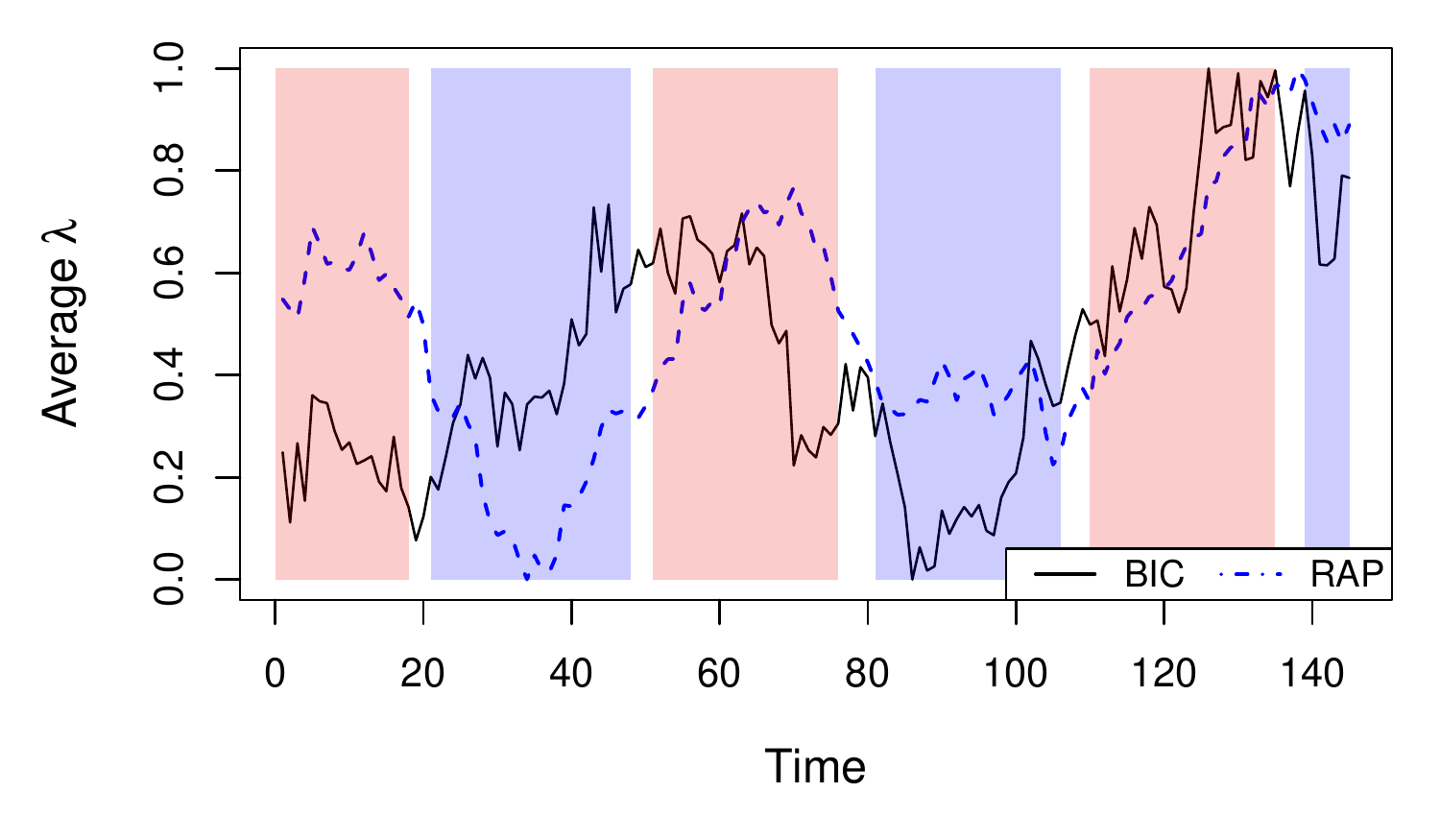}
\vspace{-3mm}
\caption{Standardized series of average $\lambda$ in the fMRI dataset. Distinct tasks are indicated
	by the background color (red indicates a neutral task, blue indicates an emotion task and white denotes the resting period).
}
\label{fig:neuro}
\href{https://github.com/QuantLet/TimeVaryingPenalization/tree/master/TVRPfmri}{\protect\quantnetright{TVRPfmri}}
\end{figure}

The modelling procedure employed for both of the datasets consisted in regressing each of the components of the multivariate time series on the rest. This way we got either 100 or 15 sequences of the Lasso parameter values, for financial and neuroimaging data respectively, which were then averaged and normalized to the [0, 1] interval as before. The resulting time series for the US stock market data are depicted in Figure \ref{fig:finance} and for the fMRI data the graphical output can be seen in Figure \ref{fig:neuro}.

From Figure \ref{fig:finance} it is visible that the values of $\lambda$ react to the situation on the market in both of the algorithms, the standard one with the BIC as a selecting rule and the RAP. Especially pronounced is the change of the values during the financial crisis of 2008-2009 where the volatility observable on the market was elevated and thus results in increased values of the Lasso parameter, too. 
Interestingly, both of the considered methods react instantly if some change occurs, but take a different amount of observations to adjust back to the standard situation.

Figure \ref{fig:neuro} shows the time series of the average regularization parameter over eight distinct subjects performing 
an emotion related task. 
The task required participants to perform a series of trails 
presented in blocks. The trails either required them to decide which of the two faces presented on the bottom of the screen match the face at the top of the screen, or which of the two shapes presented at the bottom of the screen match the shape at the top of the screen. The former 
was considered the emotion task (and denoted in blue in Figure \ref{fig:neuro}) and the latter the 
neutral task (denoted in red in Figure \ref{fig:neuro}).
From Figure \ref{fig:neuro} we see clear changes in the estimated regularization parameter induced by changes in the underlying
cognitive task, and thus, changes in the connectedness of the brain regions. This finding is in line with the current trend in the study of the fMRI data, which is 
interested in quantifying and understanding the non-stationarity properties of such a data and how these 
relate to changes in cognitive state \citep{calhoun2014chronnectome}. 

\section{Discussion} 

In this work, we have highlighted and provided extensive empirical evidence for 
the various statistical properties which affect the optimal choice of a regularization parameter in a 
penalized linear regression model. Based on the theory of the Lasso, we specifically consider three distinct properties:
the variance of residuals,
 the $\ell_0$ and $\ell_1$ norms of the regression coefficients and the 
covariance structure of the design matrix. Throughout a series of experiments, we confirm the 
manner in which each of these properties affects the optimal choice of the regularization parameter.
We relate the dependencies between each of the aforementioned statistical properties and 
estimated regularization parameter to the theoretical properties 
presented in \cite{Osborne2000lassodual}.
In particular, we conclude that:
\begin{itemize}
	\item There is a (positive) linear relationship between changes in the 
	variance of residuals, $\sigma^2$, and the estimated regularization parameter, as clearly 
	demonstrated in Figure~\ref{fig:deltasigma}. 
	\item There is a (negative) linear relationship between changes in the 
	size of the active set (either $ell_0$ or $\ell_1$ norm) and the 
	estimated regularization parameter, as  
	shown in Figure~\ref{fig:deltaq}. 
	\item There is a non-linear relationship between changes in the correlation structure in the design matrix and the estimated regularization parameter, as visualized in Figure~\ref{fig:deltarho}. 
\end{itemize}

We further provide a series of experiments where two of the statistical properties jointly 
varied in order to demonstrate the possibility of having non-stationary time-series data
where the optimal regularization parameter does not alter. This is most clearly seen in the case of 
changes in the active set, $q$, together with changes in the residual variance, $\sigma^2$,
shown in Figure \ref{fig:QvsSigma}. 

Finally, we conclude by two case studies involving high-dimensional time-series 
data in the context of finance and neuroimaging. Both datasets demonstrate significant 
temporal variability in the estimated regularization parameter, thereby validating the 
need for the methods through which to iteratively tune such a parameter. 

In conclusion, the purpose of this letter is to highlight and rigorously catalog the 
various statistical properties which may lead to changes in the choice
of the regularization parameters in $\ell_1$ penalized models. Such models are widely 
employed, indicating that an appreciation of the relationships between 
the various statistical properties of the data and the choice of 
the regularization parameter is important. 


\bibliographystyle{plainnat}
\bibliography{library.bib}

%

\end{document}